\documentclass{ws-ijmpd}
\begin{document}

\markboth{A.A. Ivanov}
{Testing for uniformity of UHECR arrival directions}

\title{Testing for uniformity of Ultra-High Energy Cosmic Ray\\ arrival directions}

\author{A.A. Ivanov}

\address{Shafer Institute for Cosmophysical Research and Aeronomy,\\
Lenin avenue 31, Yakutsk 677980, Russia\\
ivanov@ikfia.ysn.ru}

\maketitle

\begin{abstract}
Arrival directions of ultra-high energy cosmic rays (UHECRs) exhibit mainly an isotropic distribution with some small deviations in particular energy bins. In this paper, the Yakutsk array data are tested for circular uniformity of arrival directions in right ascension using 2 methods appropriate for the energy ranges below and above $10^{18}$ eV. No statistically significant deviation from uniformity is found in the arrival directions of cosmic rays (CRs) detected within the observation period 1974--2000.
\end{abstract}

\keywords{Cosmic rays; extensive air showers; arrival directions.}


\section{Introduction}	
A widely used technique to search for large-scale anisotropies in the arrival directions of CRs is the analysis in right ascension (RA), using harmonic analysis or another convenient method. The major difficulty in this approach consists of the evaluation of the directional exposure of the experiment, which is distorted by instrumental errors and weather conditions. In the case of the Yakutsk Array data, only at energies above $1$ EeV = $10^{18}$ eV do these effects appear marginal compared to the statistical errors in the estimation of extensive air shower (EAS) parameters rising with energy.

Accordingly, we divide our target energy range into 2 parts: below and above $E_{thr}=1$ EeV, where different methods of analysis are applicable. In practice, measurement below $E_{thr}$ is complicated by the need to correct the counting rate for instrumental and atmospheric effects, which must be done to prevent the introduction of artificial variations in the CR flux.

For example, the East-West method, being based on a differential technique, was designed to avoid introducing such corrections, preventing the possible associated systematics to affect the results. The original idea\cite{EWM} was proposed to be applied in the analysis of the data from the Mt. Norikura array.

The method was analyzed in detail by Bonino et al.\cite{Bonino} using simulations reproducing realistic conditions of a ground experiment subject to artificial modulations at both the diurnal and the seasonal time scales. They explain the principle of the East-West method as: "... aimed at reconstructing the equatorial component of a genuine large scale pattern by using only the difference of the counting rates of the Eastern and Western hemispheres. The effects of experimental origin, being independent of the incoming direction, are expected to be removed through the subtraction".

In this paper, another approach is used, named the South-North method (SNM), which is based on comparison of the data samples in independent right ascension circles. The principle of the method and results of the analysis as applied to the Yakutsk Array data are given in Section 2.

Above $E_{thr}$ the array exposure can be assumed uniform in RA, and a version of harmonic analysis adapted to test for uniformity of the distribution focusing on the first harmonic phase is used here. The method is described in Section 3.

A subset of the Yakutsk Array data consisting of 590887 EAS events in the energy range $E>0.1$ EeV with zenith angles $\theta<60^0$ detected within the array area during the observation period 1974--2000 is used in the analysis. The rest of the data detected after 2000 will be analyzed later.


\section{$E<10^{18}$ eV. The South-North method}
Earth's rotation gives the possibility for the surface arrays to scan celestial sphere mapping as the right ascension distribution of CR arrival directions. Scintillation counters of the Yakutsk Array have a 24-hour duty cycle resulting in an almost uniform directional exposure.\cite{JETP,MSU} However, at energies below $10^{18}$ eV, there are deviations from the uniformity, which are caused by detector malfunctions and variations of the atmospheric conditions. Estimation of the diurnal and seasonal variations of the Yakutsk Array exposure have been given previously.\cite{Pravdin,Malfa}

In this paper, however, estimations of the array exposure and corrections are not used. Instead, the RA distribution of CRs within the declination interval is tested for uniformity. The main assumption is that there is only one separable source of CRs vs isotropic background in the data observed, if any; and it is restricted in angular size. Dividing a bulk of the data into samples within declination bins, say, $\delta\leq0$ and $\delta>0$, one can compare RA distributions using the $\chi^2$-test with statistic $\Sigma_{i=1}^n (W_i-E_i)^2/E_i$, where $W_i,E_i$ are observed and expected frequencies in $n$ intervals.

A choice of the particular goodness-of-fit test is based on the result of comparative analysis\cite{Lemeshko} of the statistical power of the most popular tests in the case of simple hypotheses $H(x)=H(x,q)$, where $q$ is a known parameter. It was shown that the Pearson $\chi^2$ test with asymptotically optimal grouping is the most powerful test in comparison to Anderson-Darling $\Omega^2$, Mises $\omega^2$, and Kolmogorov-Smirnov tests.

A distinctive feature of the method is the same instrumental and weather non-uniformities of the distributions observed in different declination bins.

\begin{figure}[t]\centering
\includegraphics[width=0.7\columnwidth]{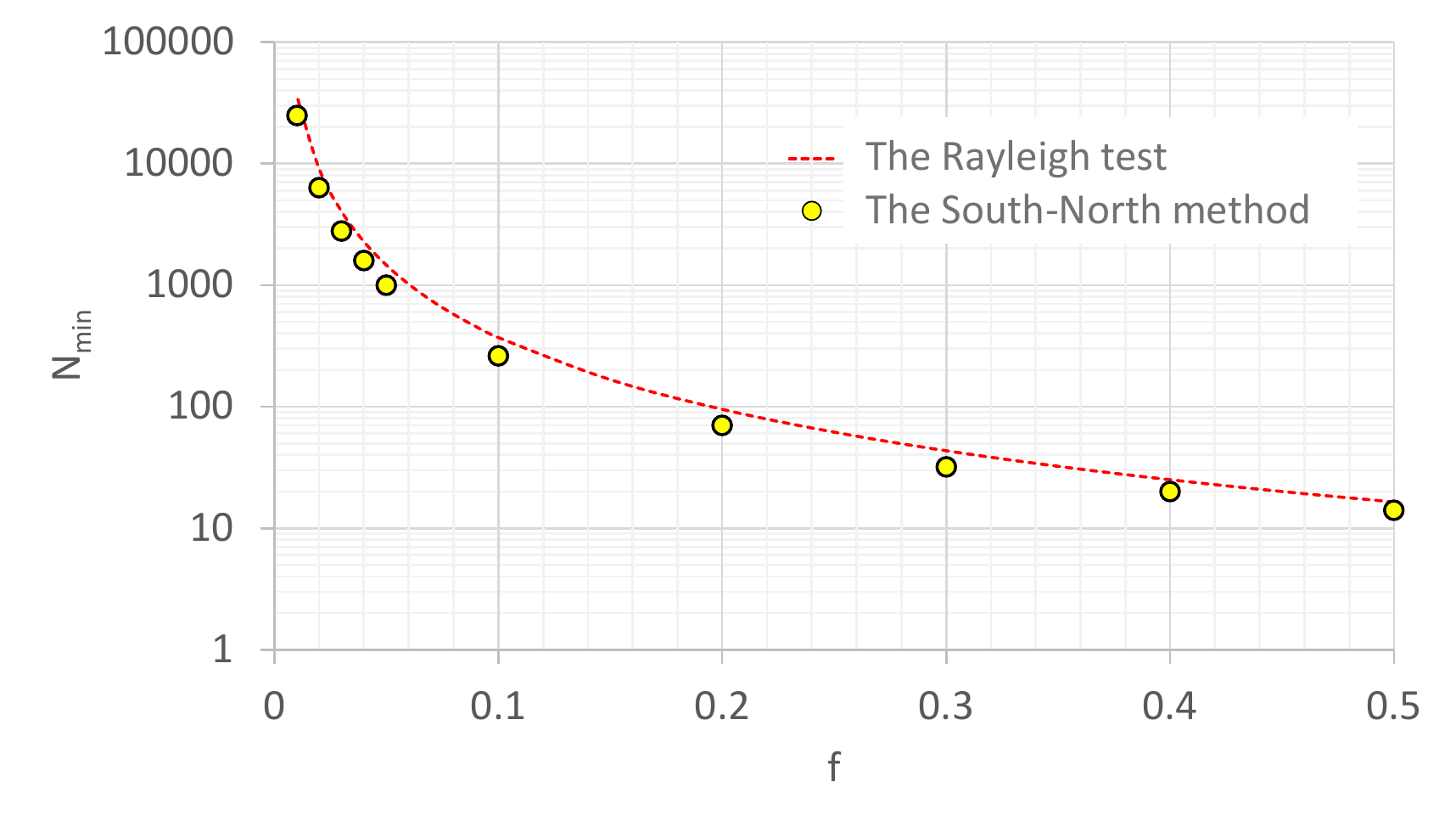}
  \caption{Statistical power of the South-North method compared to the Rayleigh test. The minimum sample size, $N_{min}$, needed to reject a null hypothesis if there is a separable source of CRs giving a fraction, $f$, of the total flux.}
\label{Fig:Power}\end{figure}

If there is no significant deviation of the samples from each other, one can conclude that: i) the distributions are uniform (taking into account the array exposure); or ii) the same source is lurking in both samples. The restricted angular size of the source excludes the second alternative, especially in the case of 3 or more samples.


\subsection{Statistical power of SNM}
The statistical power of the method is its efficiency depending on the sample size $N$. A lower limit of $N$ must be found, needed to reject the isotropic hypothesis, $H_0$, at a confidence level of 99\% when an alternative hypothesis, $H_1$, is true. To estimate $N_{min}$, $H_1$ was used consisting of a separate CR source as a $\delta$-function located in $\alpha_0$, yielding the fraction $f$ of the total CR flux, and all other sources forming an isotropic background that provide $(1-f)$ of the flux. The second sample consists of the isotropic CRs.

A Monte Carlo algorithm is used to model $H_0$ and $H_1$ with the sample sizes provided by the Yakutsk Array data. The result of the SNM simulation is illustrated in Fig. \ref{Fig:Power} compared to the Rayleigh test.\cite{MinWidth} The former method appears to be somewhat more powerful than the latter. A sample of $N>300$ EAS events is needed in both cases, for example, to reject the null hypothesis if there is a point source providing $\sim10\%$ of CR flux.


\subsection{Application of SNM to the Yakutsk Array data}
Right ascension distributions of CR arrival directions are sampled in energy, $\Delta\lg(E)=0.25$, and declination bins, $\Delta\delta=30^0$. The energy range suitable for the South-North method is $10^{17}<E<10^{18}$ eV in the case of the Yakutsk Array data. The resultant samples of data are bounded into 4 energy and 3 declination bins. The RA interval is divided into $N_{RA}=9$ equidistant bins.

\begin{table}[t]
\tbl{Probability, $P$, of the 2 samples of RA distribution being consistent, and upper limit, $f_L$, of the fraction of CRs from a source.}
{\begin{tabular}{|c|r|r|r|r|r|r|r|r|r|}
\hline lg(E, eV) & \multicolumn{3}{c|}{Middle-North} & \multicolumn{3}{c|}{Middle-South}
       & \multicolumn{3}{c|}{South-North}\\ \cline{2-10}
 bins & N  & $P,\%$ & $f_L,\%$ & N & $P,\%$ & $f_L,\%$ & N & $P,\%$      & $f_L,\%$\\ \hline
17.00-17.25 & 61477 &   5.72 &   0.64 & 11728 &  73.43 & 1.45 & 11728 & 78.96 & 1.45 \\ \hline
17.25-17.50 & 95573 &  13.22 &   0.51 & 37576 &  20.87 & 0.82 & 37576 &  2.20 & 0.82 \\ \hline
17.50-17.75 & 26078 &  26.64 &   0.98 & 17421 &  31.04 & 1.19 & 17421 &  6.03 & 1.19 \\ \hline
17.75-18.00 & 16137 &   3.85 &   1.25 & 12855 &   1.18 & 1.39 & 12855 & 49.49 & 1.39 \\ \hline
\end{tabular}\label{Table: Probability}}
\end{table}

In a given energy interval, 3 samples of RA distribution ($North:60^0<\delta\leq90^0$, $Middle:30^0<\delta\leq60^0$, $South:0^0<\delta\leq30^0$) are compared in pairs using Pearson's $\chi^2$ goodness-of-fit test with 8 degrees of freedom.

In Table 1 the resultant probabilities, $P$, are given for the $\chi^2$ random variable to be equal or greater than the test-statistic for the pair of samples. In the case $P<1$\% 2 samples are significantly different. All pairs exhibit the probabilities above this limit.

One can conclude that 3 samples: $North, Middle, South$ in declination bins are compatible, so there is no separable source of CRs with the angular size below $30^0$ in the 4 energy intervals considered.

Upper limits of the fraction of CRs from the source under the $H_1$ hypothesis have been derived, using the statistical power of SNM applied to the pair of data samples detected with the Yakutsk Array in declination bins.

The Monte Carlo code comprises a point source in the case of $H_1$, however, for the number of equidistant intervals $N_{RA}=9$ used here to calculate the $\chi^2$-statistic, it is equivalent to the source of the width $\leq40^0$.

For a given number of events\footnote{to be definite, the minimum of 2 sample sizes}, $N$, there is a limited fraction of CRs from a separable source, indistinguishable to the statistical test. An illustration can be found in Fig. \ref{Fig:Power}. Upper limits calculated for the pairs of samples, $f_L$, in energy bins are given in Table \ref{Table: Probability}. The overall limits independent of pairs in declination bins are 1.45, 0.82, 1.19, and 1.39 \% in the 4 corresponding energy bins.


\section{$E>10^{18}$ eV. Double harmonic analysis}
Harmonic analysis in RA of CR arrival directions detected with the Yakutsk Array before 2000 had found no statistically significant deviation from an isotropic distribution.\cite{Kashiwa} However, the Telescope Array collaboration reported\cite{HotSpot} on a cluster of events, a `hotspot', found above 57~EeV in the northern sky at a significance level of 5.1$\sigma$.

Additionally, the first harmonic phase exhibits a non-uniform behavior: a gradual increase with energy in RA.\cite{WG,Anova} This finding is potentially interesting, because with a real underlying anisotropy, a consistency of the phase measurements in ordered energy intervals is indeed expected to be revealed with a smaller number of events than that needed to detect the amplitude with a high statistical significance.

While the most convenient way of testing for the uniformity of arrival directions seems to be an analysis of the minimal width of the distribution\cite{MinWidth}, here another approach is followed, motivated by the findings mentioned above, aimed at the phase behavior, namely, analysis of the first harmonic phases; in other words, double harmonic analysis of the RA distribution.

\begin{table}[t]
\tbl{Double harmonic analysis of RA distribution detected before 2000 in Yakutsk. The mean values, $\langle\phi_1\rangle$, and r.m.s. deviations of the first harmonic phases, $\delta\phi_1$, in energy bins give amplitude $A_1^\phi$ with a probability $P, \%$.}
{\begin{tabular}{|l|r|r|r|r|r|r|} \hline
lg(E, eV) bins & $N$ & $M$ & $\langle\phi_1\rangle$ & $\delta\phi_1$ & $A_1^\phi$,\% & $P, \%$ \\ \hline
18.0-18.5 & 28136  & 168 &  2.9 &  102.4 &  2.2 & 98.1 \\ \hline
18.5-19.0 &  4141  &  64 & -5.2 &  102.2 & 16.3 & 65.4 \\ \hline
19.0-19.5 &   437  &  21 & 33.2 &   76.4 & 73.3 &  5.9 \\ \hline
19.5-20.0 &    55  &   7 & 47.2 &   91.6 & 64.4 & 48.4 \\ \hline
\end{tabular}\label{Table: Double}}
\end{table}

The phase of the first harmonic points to the excess flux of CRs in RA distribution, if there is a source, or is uniformly distributed otherwise. It is natural to apply the Rayleigh test to find the possible non-uniformity of the phase. In this way, the data in the given energy interval are divided additionally into a set of $M$ sub-samples with $n$ EAS events. A set of $M$ directions formed by the phases of the first harmonics in sub-samples, in turn, gives the first harmonic amplitude, $A_1^\phi$. The Rayleigh probability $P(>A_1)=\exp(-\frac{MA_1^2}{4})$ states the isotropic amplitude to be larger than the observed $A_1^\phi$ by chance.
It can be emphasized that the only objective of our double harmonic analysis is - whether the first harmonic phases in the energy interval are uniformly distributed or not?

\begin{figure}[t]\centering
\includegraphics[width=0.7\columnwidth]{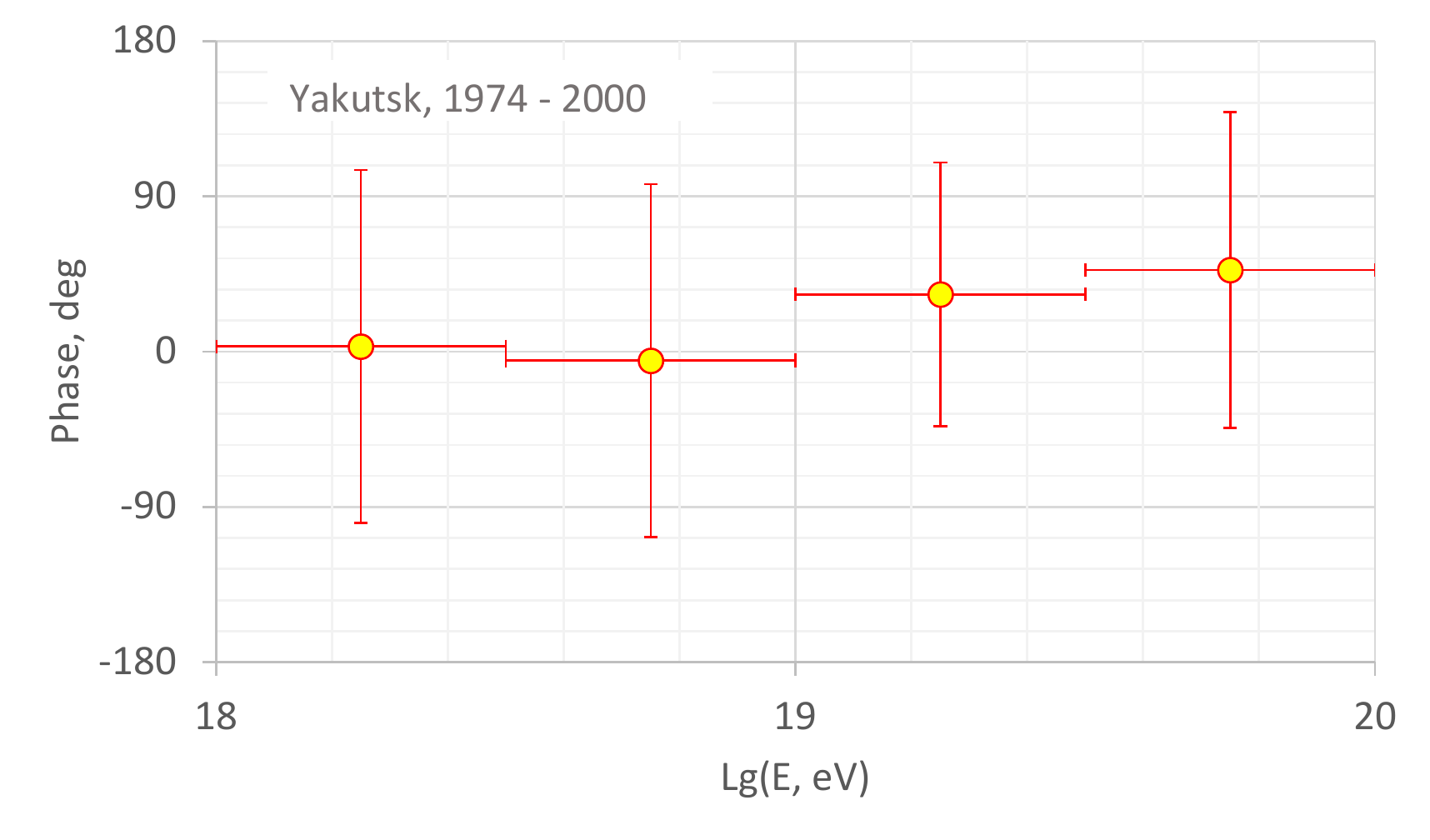}
  \caption{Phase of the first harmonic as a function of energy. Horizontal bars indicate energy bins, while vertical bars are r.m.s. deviations of phases within bins.}
\label{Fig:Phase}\end{figure}

The Yakutsk Array data consisting of CRs detected before 2000 in 4 energy intervals are divided into sub-samples of size $M=[\sqrt{N}]$, where $N$ is the number of CRs in the energy interval. The phases, $\phi_1$, of the first harmonic amplitudes of $n=[N/M]$ points in RA distribution are the input data for harmonic analysis of phases. The results are given in Table 2 and Fig. \ref{Fig:Phase}.

Rayleigh test results in a uniform distribution of phases in all energy intervals ($P>1\%$). However, the r.m.s. deviation of the phases, $\delta\phi_1$, at energies above $10^{19}$ eV is somewhat lower than the value of $103.9^0$ expected for isotropy. It has been analyzed in detail previously.\cite{Anova} The mean value of the phase $33.2^0\pm 76.4^0$ in the energy interval $10<E<31.6$ EeV is in agreement with other measurements\cite{Anova,PAOphase}, while $\langle\phi_1\rangle\sim 0^0$ below 10 EeV is an artefact of the uniform distribution averaged in the interval $\phi_1\in(-180^0,180^0)$.

In the end, upper limits are set on the fraction of CRs from a separable source under the $H_1$ hypothesis using the Rayleigh test applied to RA distribution samples of the size N from Table \ref{Table: Double}: $f<1.13;2.96;9.2;26.53\%$ in the energy bins $E\in(1,3.2);(3.2,10);(10,31.6);(31.6,100)$ EeV, respectively.


\section{Conclusions}
The Yakutsk Array data of CRs detected during the observation period 1974--2000 is used to analyze arrival directions. Two methods are applied to test for the uniformity of the right ascension distribution: the South-North method at energies below $E_{thr}=10^{18}$ eV where the diurnal and seasonal variations of the array exposure are unavoidable, and double harmonic analysis above the threshold.

At $E<E_{thr}$ 3 samples of RA distribution in different declination bins are compatible. The conclusion is that there is no single separable source of CRs with an angular size below $30^0$ detectable in the 4 energy intervals considered.

Above $E_{thr}$ the first harmonic phase exhibits the RA distribution having no statistically significant deviation from isotropy in energy intervals of the width $\Delta\lg E=0.5$. In the energy intervals considered, the upper limits are determined for the fraction of CRs from a separable source of limited angular size using statistical powers of the 2 methods.


\section*{Acknowledgements}
The author is grateful to the Yakutsk Array staff for data acquisition and analysis. The work is supported by RFBR grant no. 13-02-12036.

\end{document}